\documentclass
[preprintnumbers,noshowpacs,bibnotes,10pt,twocolumn,prb]{revtex4}%
\usepackage{graphicx}
\usepackage{amsmath}
\usepackage{amsfonts}
\usepackage{amssymb}%
\providecommand{\U}[1]{\protect\rule{.1in}{.1in}}
\providecommand{\U}[1]{\protect\rule{.1in}{.1in}}
\providecommand{\U}[1]{\protect\rule{.1in}{.1in}}
\begin{document}
\preprint{ }
\title{Self-oscillations in a superconducting stripline resonator integrated with a DC-SQUID}
\author{Eran Segev}
\email{segeve@tx.technion.ac.il}
\author{Oren Suchoi}
\author{Oleg Shtempluck}
\author{Eyal Buks}
\affiliation{Department of Electrical Engineering, Technion, Haifa 32000, Israel}
\date{\today}

\begin{abstract}
We study self-sustained oscillations (SO) in a Nb superconducting stripline
resonators (SSR) integrated with a DC superconducting quantum interface
devices (SQUID). We find that both the power threshold where these
oscillations start and the oscillations frequency are periodic in the applied
magnetic flux threading the SQUID loop. A theoretical model which attributes
the SO to a thermal instability in the DC-SQUID yields a good agreement with
the experimental results. This flux dependant nonlinearity may be used for
quantum state reading of a qubit-SSR integrated device.

\end{abstract}

\pacs{74.40.+k, 02.50.Ey, 85.25.-j}
\maketitle

We study thermal instability in superconducting stripline resonators (SSR's)
working at gigahertz frequencies. We have recently demonstrated how thermal
instability can create extremely strong nonlinearity in such resonators
\cite{segev06c,segev08a}. This nonlinearity is manifested by self-sustained
oscillations (SO) at megahertz frequencies, strong intermodulation gain,
stochastic resonance, sensitive radiation detection, and
more\cite{GilBachar08a}. In the present work we have integrated a SSR with a
DC superconducting quantum interface devices (SQUID). Similar configurations
have been recently studied by other groups and it was shown that such devices
can be used as readout and coupling elements for qubits
\cite{QubitResCoup_Wallraff04,QubitResCoup_Lupacu06,qubitResReadout_Lee07}. In
addition, it was shown that further improvement in the sensitivity of the
readout process of the qubit is possible by biasing the resonator to a state
of nonlinear responsitivity
\cite{ResCoup_Siddiqi06,ResCoup_Boaknin07,ResCoup_Metcalfe07,qubitResReadout_Lee07,QubitResCoup_Lupacu06}%
.

\begin{figure}
[ptb]
\begin{center}
\includegraphics[
height=3.1498in,
width=3.4006in
]%
{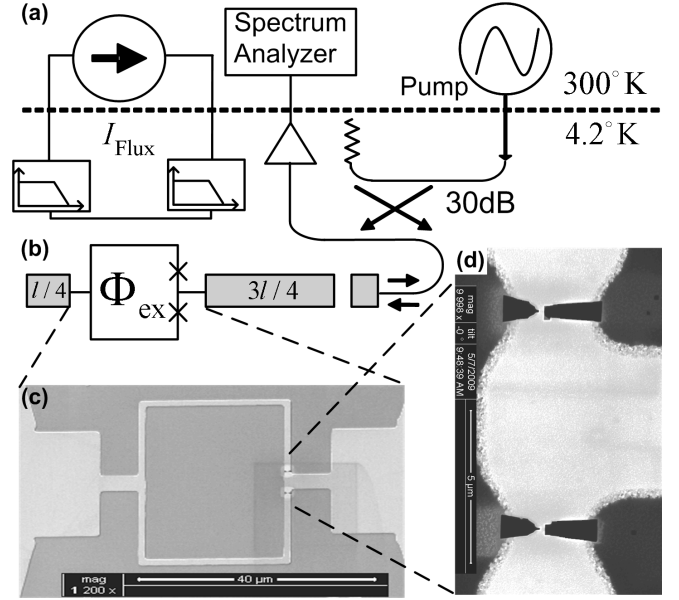}%
\caption{$\mathrm{(a)}$ Measurement setup. $\mathrm{(b)}$ Schematic layout of
our device. $\mathrm{(c)}$ SEM image of the DC-SQUID. $\mathrm{(d)}$ SEM image
of the two nano-bridge-based Josephson junctions.}%
\label{Schematics}%
\end{center}
\end{figure}

Our experiments are performed using the setup depicted in Fig.
\ref{Schematics}$(\mathrm{a})$. We study the response of the integrated SSR to
a monochromatic injected pump tone that drives one of the resonance modes, and
measure the reflected power spectrum by a spectrum analyzer. We find that
there is a certain range in the plane of the pump-frequency pump-power
parameters, in which intrinsic SO occur in the resonator. These oscillations
are manifested by the appearance of sidebands in the reflected power spectrum.
In addition, we apply bias magnetic flux through the DC-SQUID and find that
both the threshold where these oscillations start and their frequency are
periodic in the applied magnetic flux, having a periodicity of one flux
quantum. We extend our theoretical model \cite{segev06c}, which has originally
considered SO and thermal instability in a SSR integrated with a single
micro-bridge, to include this flux dependency and find a good agreement with
the experimental results.%

A simplified circuit layout of our device is illustrated in Fig.
\ref{Schematics}$(\mathrm{b})$. We fabricate our devices on a Silicon wafer,
covered by a thin layer of Silicon Nitride. Each device is made of a thin
layer $(<100%
\operatorname{nm}%
)$ of Niobium and composed of a stripline resonator having a
DC-SQUID\ (\ref{Schematics}$(\mathrm{c}))$ monolithically embedded into its
structure. The resonator length is $l=18%
\operatorname{mm}%
$, and its first resonance mode is found at $f_{1}=\omega_{1}/2\pi=3.006%
\operatorname{GHz}%
$. The DC-SQUID has two nano-bridges (\ref{Schematics}$(\mathrm{d}))$, one in
each of its two arms. Their size is typically $100%
\operatorname{nm}%
^{2}$ and therefore each nano-bridge functions as a weak-link that
approximately can be regarded as a regular Josephson junction
\cite{Troeman_024509,Troeman_2152_2008}. A feed-line, weakly coupled to the
resonator, is employed for delivering the input and output signals. An on-chip
filtered DC bias line passing near the DC-SQUID is used to apply magnetic flux
through the SQUID. Some measurements are carried out while the device is fully
immersed in liquid Helium, while others in a dilution refrigerator where the
device is in vacuum. Further design considerations and fabrication details can
be found elsewhere \cite{Oren09a}.%

\begin{figure}
[ptb]
\begin{center}
\includegraphics[
height=2.5255in,
width=3.3723in
]%
{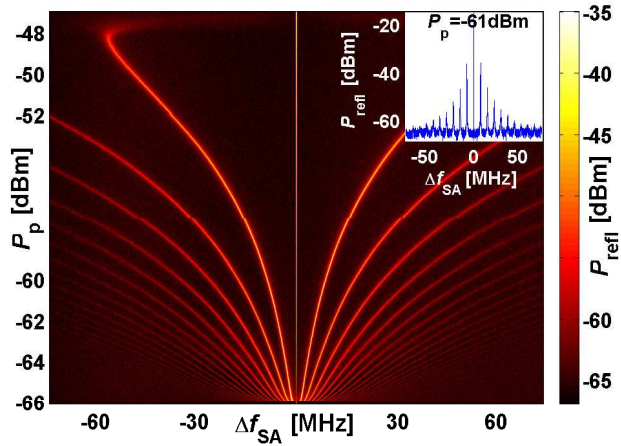}%
\caption{Typical experimental results of the SO phenomenon. The color-map
shows the reflected power $P_{\mathrm{refl}}$ as a function of the input pump
power $P_{\mathrm{p}}$ and the measured frequency $f_{\mathrm{SA}}$, centered
on the pump frequency, which coincides with the first resonance frequency
$f_{1}$ ($\Delta f_{SA}=f_{SA}-f_{1}$). Note that the data is truncated at
$P_{\mathrm{refl}}=-35~\mathrm{dBm}$ for clarity. The inset shows a cut of
that measurement obtained with $P_{\mathrm{p}}=-61\mathrm{dBm}$.}%
\label{typicalSO}%
\end{center}
\end{figure}

Figure 2 exhibits SO in the power reflected off the resonator
\cite{segev06b,segev06c}. In this measurement we inject into the resonator an
input pump tone having a monochromatic frequency $\omega_{\mathrm{p}}$ $=$
$\omega_{1}$, and measure the reflected power spectrum around $\omega_{1}$
while varying the input pump power $P_{\mathrm{p}}$. At relatively low
$\left(  P_{\mathrm{p}}\lesssim-66\mathrm{dBm}\right)  $ and high $\left(
P_{\mathrm{p}}\gtrsim-48\mathrm{dBm}\right)  $ pump powers the response of the
resonator is linear, namely, the reflected power spectrum contains a single
spectral component at the frequency of the stimulating pump tone. In between
these two power thresholds the resonator self oscillates and regular
modulation of the pump tone occurs. As a result, the reflected spectrum
contains several orders of modulation products realized by rather strong and
sharp sidebands (see inset of Fig. \ref{typicalSO}) that extend to both sides
of the pump tone frequency. The SO frequency, defined as the frequency
difference between the pump frequency and the primary sideband, ranges between
few to tens of megahertz and increases with the pump power.%

\begin{figure}
[ptb]
\begin{center}
\includegraphics[
height=2.5645in,
width=3.4006in
]%
{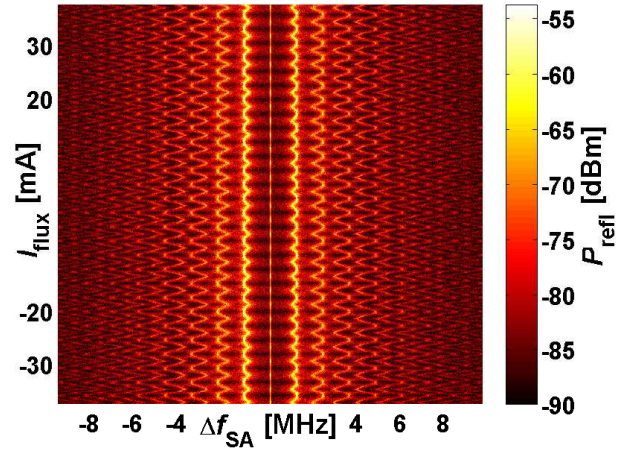}%
\caption{Typical experimental results of flux dependent SO. The color-map
shows the reflected power $P_{\mathrm{refl}}$ as a function of the bias DC
current $I_{\mathrm{flux}}$ and the measured frequency $f_{\mathrm{SA}}$
centered on the pump frequency $f_{\mathrm{p}}=f_{1}$ $(\Delta f_{\mathrm{SA}%
}=f_{\mathrm{SA}}-f_{\mathrm{1}})$.}%
\label{SMvsFlux3D}%
\end{center}
\end{figure}

\begin{figure}
[ptb]
\begin{center}
\includegraphics[
height=2.5645in,
width=3.4006in
]%
{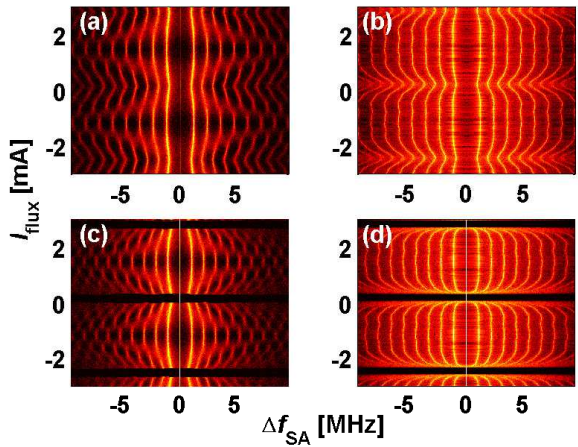}%
\caption{Typical experimental and numerical results of flux dependent SO. Each
panel shows a color-map of the reflected power $P_{\mathrm{refl}}$ as a
function of the bias DC current $I_{\mathrm{flux}}$ and the measured frequency
$f_{\mathrm{SA}}$ centered on the pump frequency $f_{\mathrm{p}}=f_{1}$
$(\Delta f_{\mathrm{SA}}=f_{\mathrm{SA}}-f_{\mathrm{1}})$. Panels
$(\mathrm{a})$ and $(\mathrm{c})$ show experimental results obtained at pump
power above and equals the low threshold power, respectively. Panels
$(\mathrm{b})$ and $(\mathrm{d})$ show the corresponding theoretical results
obtained by numerically integrating the equations of motion.}%
\label{SMvsFlux3DExpAndSim}%
\end{center}
\end{figure}

The dependence of the SO on the applied magnetic flux is shown in Fig.
\ref{SMvsFlux3D}. In this measurement we inject a pump tone having a
stationary frequency $\left(  \omega_{\mathrm{p}}\simeq\omega_{1}\right)  $
and power $\left(  P_{\mathrm{p}}=-51.5\mathrm{dBm}\right)  $, and measure the
reflected power spectrum around $\omega_{1}$ while varying the DC bias
current. As shown, the SO frequency is periodically changed by the applied
magnetic flux. The periodicity is one flux quantum and the relative change is
typically about $20\%$. Figures \ref{SMvsFlux3DExpAndSim}$\mathrm{(a)}$ and
$\mathrm{(c)}$ show two measurements of flux dependent SO obtained with pump
power above and equal the threshold power $\left(  P_{\mathrm{p,(a)}%
}=-51.5\mathrm{dBm}\text{, }P_{\mathrm{p,(c)}}=-51.6\mathrm{dBm}\right)  $,
respectively. The later measurement demonstrates how magnetic flux can switch
the resonator from a steady state response to a limit-cycle state where it
experiences SO.

To account for our results we model our device as a transmission line
resonator interrupted by a DC-SQUID. The impedance of the DC-SQUID is composed
of a constant inductor in series with a flux-dependent inductor shunted by a
parallel resistor \cite{Oren09a}. The flux dependence of the DC-SQUID
impedance gives rise to periodic dependence of the SSR's damping rates on the
applied magnetic field (the change in the SSR resonance frequency is
relatively small).

The dynamics of our system can be captured thus by two coupled equations of
motion \cite{segev06c}. The first equation describes the dynamics of a
resonator driven by feed-line carrying an incident coherent tone
$b^{\mathrm{in}}=b_{0}^{\mathrm{in}}e^{-i\omega_{\mathrm{p}}t}$, where
$b_{0}^{\mathrm{in}}$ is constant complex amplitude and $\omega_{\mathrm{p}%
}=2\pi f_{\mathrm{p}}$ is the driving angular frequency.\ The mode amplitude
inside the resonator can be written as $Be^{-i\omega_{\mathrm{p}}t}$, where
$B\left(  t\right)  $ is complex amplitude, which is assumed to vary slowly on
a time scale of $1/\omega_{\mathrm{p}}$. \ In this approximation, assuming a
noiseless system, the equation of motion of $B$ reads \cite{Squeezing_Yurke05}%
\begin{equation}
\frac{\mathrm{d}B}{\mathrm{d}t}=\left[  i\left(  \omega_{\mathrm{p}}%
-\omega_{0}\right)  -\gamma\right]  B-i\sqrt{2\gamma_{1}}b^{\mathrm{in}},
\label{dB/dt}%
\end{equation}
where $\omega_{0}$ is the angular resonance frequency and $\gamma\left(
T\right)  =\gamma_{1}+\gamma_{2}\left(  T,\Phi_{ex}\right)  $, where
$\gamma_{1}$ is the coupling coefficient between the resonator and the
feed-line and $\gamma_{2}\left(  T,\Phi_{ex}\right)  $ is the temperature and
flux dependent damping rate of the mode.

The heat-balance equation for the temperature $T$ of the nano-bridges
composing the DC-SQUID is given by \cite{segev06b}
\begin{equation}
C\frac{\mathrm{d}T}{\mathrm{d}t}=2\hslash\omega_{0}\gamma_{2}\left\vert
B\right\vert ^{2}-H\left(  T-T_{0}\right)  , \label{dT/dt}%
\end{equation}
where $C$ is the thermal heat capacity, $H$ is the heat transfer coefficient,
and $T_{0}$ is the temperature of the coolant.

Coupling between Eqs. (\ref{dB/dt}) and (\ref{dT/dt}) originates by the
dependence of the damping rate $\gamma_{2}\left(  T,\Phi_{ex}\right)  $ of the
driven mode on the impedance of the DC-SQUID \cite{supRes_Saeedkia05}, which
in turn depends on the temperature of its nano-bridges and on the applied
magnetic flux. We assume a simple case, where the dependence on the
temperature $T$ is described by a step function that occurs at the critical
temperature $T_{\mathrm{c}}$ of the superconductor. We further assume that
only when the nano-bridges are in a superconducting phase the damping rate
depends on the external flux. Namely, $\gamma_{2}$ takes a flux dependent
value $\gamma_{2\mathrm{s}}\left(  \Phi_{ex}\right)  $ when the nano-bridges
are in a superconducting phase $(T<T_{\mathrm{c}})$ and a flux-independent
value $\gamma_{2\mathrm{n}}$ when they are is a normal-conducting phase
$(T>T_{\mathrm{c}})$.%

The results of a numerical integration of the equations of motion are shown in
Fig. \ref{SMvsFlux3DExpAndSim}$\mathrm{(b)}$ and $\mathrm{(d)}$. Our model
qualitatively reproduces the same flux dependency of the SM frequency on the
magnetic flux. The parameters used for the numerical simulation were obtained
as follows. The thermal heat capacity $C=4.5\operatorname{nJ}\operatorname{cm}%
^{-2}\operatorname{K}^{-1}$ and the heat transfer coefficient\ $H=24.5%
\operatorname{mW}%
\operatorname{cm}^{-2}\operatorname{K}^{-1}$ were calculated analytically
according to Refs. \cite{kinInd_Johnson96,HED_Weiser81}. The values of the
resonance frequency and the various damping rates, $\gamma_{1\mathrm{s}%
}=2.4\operatorname{MHz}$, $\gamma_{2\mathrm{s}}\in\left[  2.93,3.32\right]  $
$\operatorname{MHz}$, $\gamma_{2\mathrm{n}}=13\operatorname{MHz}$, and
$\omega_{0}=2\pi\cdot3.006%
\operatorname{GHz}%
$ were extracted from S11 reflection coefficient measurements according to
\cite{Segev06a}.

In conclusion, we report on periodic flux-dependency of SO in a SSR integrated
with a DC-SQUID. The flux significantly modulates the oscillation frequency
and can be used to turn them on and off if the device is driven near the power
threshold. A theoretical model which attributes this behavior to thermal
instability in the DC-SQUID exhibits a good quantitative agreement with the
experimental results.

We thank Steve Shaw for valuable discussions and helpful comments. E.S. is
supported by the Adams Fellowship Program of the Israel Academy of Sciences
and Humanities. This work is supported by the German Israel Foundation under
grant 1-2038.1114.07, the Israel Science Foundation under grant 1380021, the
Deborah Foundation, the Poznanski Foundation, Russell Berrie nanotechnology
institute, and MAFAT.

\bibliographystyle{apsrev}
\bibliography{Bibilography}

\end{document}